\documentclass{SciPost}

\hypersetup{
	colorlinks,
	linkcolor={red!50!black},
	citecolor={blue!50!black},
	urlcolor={blue!80!black}
}

\usepackage[bitstream-charter]{mathdesign}
\usepackage{verbatim}

\DeclareSymbolFont{usualmathcal}{OMS}{cmsy}{m}{n}
\DeclareSymbolFontAlphabet{\mathcal}{usualmathcal}

\fancypagestyle{SPstyle}{
	\fancyhf{}
	\lhead{\colorbox{scipostblue}{\bf \color{white} ~SciPost Physics }}
	\rhead{{\bf \color{scipostdeepblue} ~Submission }}
	
	\fancyfoot[C]{\textbf{\thepage}}
}

\begin{document}
	
	\pagestyle{SPstyle}
	
	\begin{center}{\Large \textbf{\color{scipostdeepblue}{
					Large deviations in quantum dynamics and complexity
					\\
	}}}\end{center}
	
	\begin{center}\textbf{
			Xiangyu Cao \textsuperscript{1$\star$}, and
			Jorge Kurchan \textsuperscript{1$\dagger$} 
	}\end{center}
	
	\begin{center}
		{\bf 1} Laboratoire de Physique de l'\'Ecole normale sup\'erieure, ENS, Universit\'e PSL, CNRS, Sorbonne Universit\'e, Universit\'e Paris Cit\'e, F-75005 Paris, France
		
		$\star$ \href{mailto:email1}{\small xiangyu.cao@phys.ens.fr}\,,\quad
		$\dagger$ \href{mailto:email2}{\small kurchan.jorge@gmail.com}
	\end{center}

	\section*{\color{scipostdeepblue}{Abstract}}
	\textbf{\boldmath{%
			We study three definitions of large deviation in many-body quantum dynamics: (i) via the full distribution of an extensive observable, (ii) via the distribution of measurement outcomes (from a continuous monitoring of the observable) over a time interval $t \le t_{\max}$, and (iii) via the distribution of expectation values over $t \le t_{\max}$. In generic systems without conservation laws, the large deviation function (i) reaches its longtime limit at $t \sim \mathcal{O}(1)$, independently of system size $N$. (ii) and (iii) reach their longtime limit at $t \sim e^{ N}$ and $t \sim \exp(e^{N})$, respectively. Before that, there is a {\it sharp} frontier between the explored and unexplored outcomes/expectation values; their distribution equals the longtime limit truncated at values that drift with $t_{\max}$. We propose that the evolution of these values with $t_{\max}$ provides a measure of quantum complexity.    
	}}
	
	\vspace{\baselineskip}
	
	\noindent\textcolor{white!90!black}{%
		\fbox{\parbox{0.975\linewidth}{%
				\textcolor{white!40!black}{\begin{tabular}{lr}%
						\begin{minipage}{0.6\textwidth}%
							{\small Copyright attribution to authors. \newline
								This work is a submission to SciPost Physics. \newline
								License information to appear upon publication. \newline
								Publication information to appear upon publication.}
						\end{minipage} & \begin{minipage}{0.4\textwidth}
							{\small Received Date \newline Accepted Date \newline Published Date}%
						\end{minipage}
				\end{tabular}}
		}}
	}

	
	\vspace{10pt}
	\noindent\rule{\textwidth}{1pt}
	\tableofcontents
	\noindent\rule{\textwidth}{1pt}
	\vspace{10pt}

	\section{Introduction }
	\label{sec:intro}
	
	In many-body quantum systems, unitary evolution generically leads to a relaxation to thermal equilibrium. This relaxation can be characterized by the convergence of local quantities' expectation value, the decay of correlation functions, or, more information-theoretically, by the growth and saturation of entanglement entropy and out-of-time-order correlations. In a large finite system, the relaxation time, as defined by the above probes, scales at most polynomially with the system size. 
	
	However, it is believed that even long after the relaxation, the dynamics continues to generate finer and finer structures in the wavefunction and in the unitary evolution operator. To measure these fine structures requires the introduction of a quantity,  {\it complexity}, that would describe how the space of wavefunctions or unitaries is covered in greater detail by the time evolution. One motivation for complexity is the physics of black holes~\cite{Susskind2020ThreeLectures}, in particular, in the context of holography correspondence. The horizon geometry of a black hole equilibrates after a scrambling time of order
	$t_* \sim  \ln N$ ($N$ is the black hole entropy, proportional to the boundary system size), and from the outside, the black hole already appears thermal. Yet the inside of the black hole is expected to continue growing for exponentially longer times. Susskind~\cite{Susskind2020ThreeLectures} proposed that this slow interior growth is dual to the increase of quantum complexity in the boundary quantum system. 
	
	In Susskind's proposal, the quantum complexity is akin to that of Nielsen~\cite{nielsen06-1,nielsen06}, and defined as the minimal number of elementary operations required to approximate the-evolved state or the unitary operator to within a prescribed precision $\epsilon$. It has been argued that the complexity growth continues until a saturation time that scales exponentially with the system size,
	\begin{equation} \label{eq:tplateau}
		t_{\mathrm{plateau}} \sim e^{N} \log \left(\frac{1}{\epsilon}\right)
	\end{equation}
	at which point it reaches a plateau value
	\begin{equation}\label{eq:Cmax}
		\mathcal C_{\max}\sim e^{N} \log \left(\frac{1}{\epsilon}\right). 
	\end{equation}
	Thus, unlike entropy, which saturates rapidly after equilibration, complexity is expected to keep increasing for exponentially long times.  Recurrences take place at double exponential times $t \sim \exp(e^{N})$, where complexity becomes temporarily small again.  
	Another strategy of defining complexity is to use the Krylov basis~\cite{uogh,Rabinovici:2020ryf,vijay,nandy-review,Rabinovici:2025otw}, generated by an initial wavefunction (or operator), and the Hamiltonian (or the Liouvillian, respectively). In general, the wavefunction or operator delocalizes in the Krylov basis until $t \sim e^{N}$. Relations between the two complexities have been explored in Ref.~\cite{krylov-nielsen}. 
	
	In this paper, we explore alternative definitions quantum complexity, using large deviations. The basic idea is that the time evolution of many-body systems makes brief incursions into increasingly atypical regions. In (deterministic) classical many-body systems, such rare incursions are known to take place at an exponential time scale $t \sim e^{N}$, see Section~\ref{sec:classical}. Meanwhile, quantum large deviations and their time scales have not been systematically studied. In fact, as we shall show both analytically and numerically in Section~\ref{sec:quantum}, there are several large deviation time scales, depending on how the large deviation is defined: 
	\begin{itemize}
		\item The large deviation function defined by the full distribution of an extensive observable $A$, $\left< \psi |  \delta(A(t) - N a) | \psi \right>$, 
		where $\psi$ is some non-equilibrium initial state. In absence of conservation laws, this large deviation function generically equilibrates (that is, reaches its long time limit) at $t \sim 1$, independently of system size. 
		\item Under the continuous monitoring of an extensive observable, the distribution of the outcomes in a time interval $t \le t_{\max}$ also has a large deviation form, and equilibrates at $t_{\max} \sim e^{N}$ in a system of size $N$. 
		\item The distribution of the expectation value of an operator in a time interval $t \le t_{\max}$ equilibrates at $t_{\max} \sim \exp(e^{N})$. This time scale is that of Hilbert space recurrence~\cite{Venuti2015Recurrence,Kotowski2026Recurrence,GuptaShort2026Recurrence}, and also the large deviation time scale of the return amplitude, spectral form factor, and correlation functions. 
	\end{itemize}
	In the two latter cases, before the equilibration time, the distribution of outcome or expectation values have \textit{sharp} cutoffs at threshold values that slowly evolve as $t_{\max}$ increases. In this sense, these large deviations describe the quantum evolution (either intrinsic or monitored) making brief incursions into increasingly atypical regions of the Hilbert space, which we propose as a measure of complexity.

	\section{Classical large deviations} \label{sec:classical}
	We first briefly recall the behavior of large deviations of an extensive quantity in a classical many-body system. For example, in a gas of $N$ interacting particles inside a box under Hamiltonian evolution, an extensive quantity $B$ is the number of particles in the left half of the box. Suppose that we observe $B$ during $0 \le t \le t_{\max}$ and construct a histogram of its values. In the $t_{\max} \to \infty$ limit, we expect the distribution to converge to a large deviation form,
	\begin{equation}
		P_{\infty}(B) \sim \exp\left(- N f(B / N) \right),
	\end{equation}
	where the large deviation rate $f(b)$ is independent of $N$ in the thermodynamic limit $N \to \infty$. 
	
	What about the distribution for large but finite $t_{\max}$? To understand the generic form, one may consider a rare region of the phase space where $B$ takes an atypical value, with $f > 0$, and $P_\infty(B) \sim e^{- N f} \ll 1$. The dynamics will visit such regions once every $ e^{N f} $ unit of time (and, necessarily, remain there briefly). Thus, if $\tau \ll e^{N f} $, these regions will not be visited; if $\tau \gg e^{Nf}$, they will be visited many times so that the $\tau = \infty$ statistics is established. We therefore expect that,
	\begin{equation} \label{eq:Ptau}
		P_{\tau}(B) \sim 
		\begin{cases}
			\exp\left(- N f(B / N) \right) &   f(B / N) \lesssim f_*(t_{\max})   \\ 
			0 & f(B / N) \gtrsim f_* (t_{\max}) 
		\end{cases}, \quad
		\text{where }    f_* := N^{-1} \ln t_{\max}  .
	\end{equation}
	In other words, the sampling time enters through cutoffs in the large deviation distribution. When $f$ has a maximum $f_{\max}$, the finite-time distribution will converge to infinite-time one at $t_{\max} \sim e^{N f_{\max}} $, that is, when the rarest values have occurred. Eq.~\eqref{eq:Ptau} needs to be slightly amended if the initial condition has an atypical value of $B$. In that case, $B$ will quickly relax to its typical value, and the distribution after that initial transient still satisfies \eqref{eq:Ptau}. 
	
	The above argument about rare values is standard, and it underlies the random energy model \cite{derrida1981random} transition, and in general, extreme-value statistics \cite{bouchaud1997universality}. We will use it again in Section~\ref{sec:measure} and Section~\ref{sec:expectation} below. 
	
	Recently, Levesque and Sourlas~\cite{sourlas} explicitly tested the above scenario by extensive numerical simulation of a system of $N$ particles with Lennard-Jones interaction. The gas particles are initialized in a half of the box. They showed that the time it takes for $3/4$ of the particles to return to the half-box scales as $e^{N}$ (where the constant $c$ can be related to the entropy density).

	

	
	\section{Quantum large deviations}\label{sec:quantum}
	How does one generalize the notion of large deviation to the quantum realm? In this Section we shall study three possible definitions, using analytical arguments and numerical tests. The latter will be performed on the 1D kicked Ising chain of length $N$. Its discrete-time evolution is defined by the unitary evolution operator 
	\begin{align}  \label{eq:kicked-def}
		&  U = e^{-i H_z / 2} e^{-i H_x} e^{- i H_z / 2},  \\ 
		&  H_z := - \sum_{j = 1}^{N-1} Z_j Z_{j+1} (1 + 0.3 \cos(j)) - \sum_{j=1}^N Z_j,  \quad H_x := - \sum_{j=1}^{N} X_j, 
	\end{align}
	where $Z_j$ and $X_j$ are Pauli matrices acting on the $j$-th qubit. This is a chaotic quantum many-body system with no local conserved quantities and no small parameters. Any initial condition is expected to rapidly relax to the infinite-temperature state. We will use the total magnetization 
	\begin{equation}
		\mathcal{M} = \sum_{j = 1}^N Z_j 
	\end{equation}
	as an example of extensive quantity. 
	
	\subsection{Full distribution}\label{sec:full}
	We first consider the full distribution $ \mathbf{P}(\mathcal{M} = m)$ of the extensive quantity with respect to a time-evolved state $| \psi(t) \rangle = U^t | \psi \rangle$, which is also characterized by the generating function, 
	\begin{equation}
		\left< e^{\lambda \mathcal{M}} \right>_t := \langle \psi(t) | e^{\lambda \mathcal{M}} | \psi(t) \rangle = \sum_{m} \mathbf{P}(\mathcal{M} = m)  e^{\lambda m}. 
	\end{equation}
	The full distribution is akin to the two-time full counting statistics~\cite{LevitovLesovik1993,LevitovLeeLesovik1996,Klich2003}, which corresponds to the generating function $\mathrm{Tr}[ \rho e^{\lambda \mathcal{M}(t)}  e^{-\lambda \mathcal{M}}  ]$. 
	
	If $N$ is large, and when $| \psi(t) \rangle \langle \psi(t) |$ is well-approximated by the infinite-temperature state, the full distribution has a large deviation form (that follows from the binomial distribution)
	\begin{equation} \label{eq:ldf-short}
		- N^{-1} \ln \mathbf{P}(\mathcal{M} = m N) \to f(m) = \frac{1 + m}{2} \ln (1 + m) + \frac{1 - m}{2} \ln  (1 - m) , 
	\end{equation}
	and correspondingly, the generating function has the following limit,
	\begin{equation} \label{eq:gf-prediction}
		N^{-1} \ln  \left< e^{\lambda \mathcal{M}} \right>_t  \to \ln \cosh(\lambda). 
	\end{equation}

\begin{figure}
		\centering
		\includegraphics[width=1\linewidth]{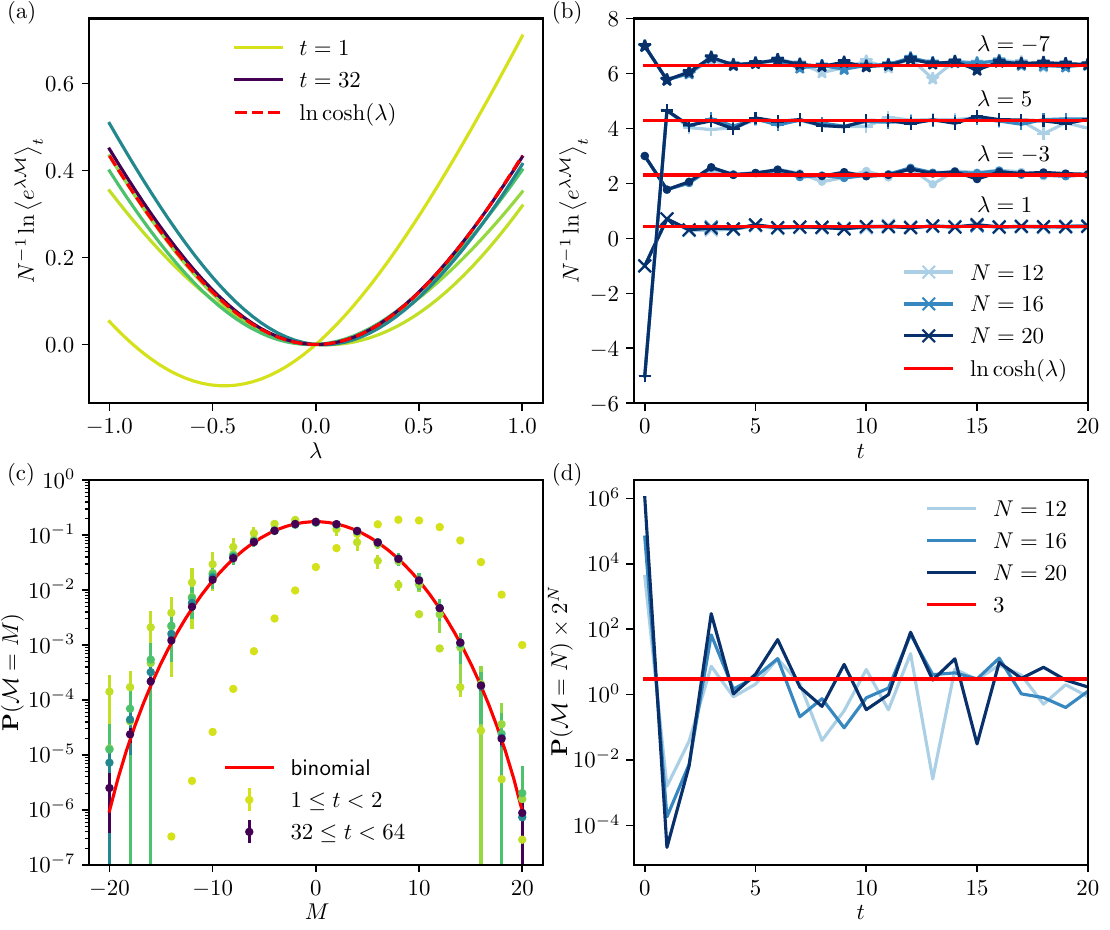}
		\caption{Full distribution and generating function of the total magnetization $\mathcal{M}$ in the kicked Ising model~\eqref{eq:kicked-def} in a quench dynamics starting from $ | \psi(t=0)\rangle= | \uparrow \uparrow \dots \uparrow \rangle $. 
			(a) the rescaled generating function $N^{-1} \left< e^{\lambda \mathcal{M}} \right>_t =N^{-1} \left< \psi(t) |e^{\lambda \mathcal{M}} | \psi(t) \right> $ at $t = 1, 2, 4, 8, 16, 32$, compared to the infinite-temperature ensemble prediction $\ln \cosh(\lambda)$. System size $N = 20$. 
			(b) Time dependence of the rescaled generating function for select values of $\lambda$ and different system sizes. The convergence to the infinite-temperature ensemble value takes place at $t \sim \mathcal{O}(1)$, independently of the system size. 
			(c) Distribution of total magnetization for $N = 20$ averaged over time windows $1 \le t < 2$, $2 \le t < 4$, \dots, $32 \le t < 64$ (the error bar is the standard deviation over time), compared to the binomial distribution.
			(d) Rescaled probability of maximum magnetization $\mathcal{M} = N$ as a function of time for $N = 12, 16, 20$, compared to the $ N \to \infty$ analytical prediction~\eqref{eq:Pinfty-prediction}.
		}    
		\label{fig:shorttime}
	\end{figure}

	When does the full distribution converge to the infinite-temperature large deviation form, including its rare event tails? (We expect a rapid relaxation of the typical-value distribution, or the generating function with $\lambda$ close to $0$.) We address this question numerically, with the initial condition $ | \psi \rangle = | \uparrow \uparrow \dots \uparrow \rangle $ with $\mathcal{M} = N$. We find that the entire distribution equilibrates at $t \sim \mathcal{O}(1)$ time, independently of the system size $N$. In Figure~\ref{fig:shorttime}, we observe that the generating function converges to the infinite-temperature value for all values of $\lambda$ at the same time. The full distribution $ \mathbf{P}(\mathcal{M} = M) $ also converges to the binomial large deviation form at $t \sim \mathcal{O}(1)$, with no time-dependent cutoff, unlike in the classical large deviation reviewed above. 
	
	In particular, the right extreme of the distribution coincides with the return probability, 
	\begin{equation}
		\mathbf{P}(\mathcal{M} = N) =  | \left< \psi | \psi(t) \right> |^2, 
	\end{equation}
	and its longtime average is given by the diagonal ensemble 
	\begin{equation}
		\mathbf{P}_\infty = \frac1{T} \sum_{t = 1}^T  | \left< \psi | \psi(t) \right> |^2 \to \sum_i | \langle \epsilon_i | \psi \rangle |^4
	\end{equation}
	where $| \epsilon_i \rangle$ are the eigenstates of $U$ (pseudo-energy eigenstates). As the kicked Ising model has time reversal symmetry, we expect that $\langle \epsilon_i | \psi \rangle$ to be distributed as \textit{real} Gaussian random variables with variance $2^{-N}$ in the $N\to\infty$ limit (we checked this numerically). Therefore, we have, from Wick theorem~\footnote{Eq.~\eqref{eq:Pinfty-prediction} means that, if one make a one-shot measurement of $\mathcal{M}$ at a large enough $t$, there is $3 \times 2^{-N}$ probability to find an outcome $N$. In Section~\ref{sec:measure}, we will see that, under \textit{repeated} measurements, this probability becomes $2^{-N}$ at late time. There is no contradiction between the two facts.},
	\begin{equation} \label{eq:Pinfty-prediction}
		2^N \mathbf{P}_\infty \to 3 \quad, N \to \infty. 
	\end{equation}
	This prediction, which goes beyond the large deviation rate~\ref{eq:ldf-short}, is compared to the finite-size, finite-time data in Figure~\ref{fig:shorttime}-(d), where we see a quantitative agreement established at an size-independent time. 
	
	We expect the equilibration at $t \sim \mathcal{O}(1)$ of the full-distribution large deviation function in absence of any conserved quantity. When the latter is present, and if the observable is correlated to its spatial distribution, the equilibration time scale will be dictated by transport, and scales with the linear system size as $t \sim L^{z}$ where $z$ is the dynamical exponent ($z= 2$ for diffusion). Note that this is still much shorter than an exponential time scale $e^{N}$ for any $ c > 0$.
	
	The above observation is in apparent disagreement with the classical intuition that it takes exponentially long time for the dynamics to stumble upon rare configurations. There is no genuine contradiction, since it is still exponentially unlikely to actually observe a rare outcome at short time. Also, the classical-mechanics picture of a single trajectory in the configuration space cannot describe the dynamics in our quantum spin chain that is far away from semiclassical limit. Instead, quantum parallelism visits to rare and typical configurations all at once. To illustrate, consider starting the dynamics with the N\'eel state $| \uparrow \downarrow \uparrow\downarrow \dots \rangle $ which has a typical magnetization value. Then, a single application of $e^{ - i \theta \sum_j X_j } $, which is contained in the Floquet unitary, gives 
	\begin{align}
		&	e^{- i \theta \sum_j X_j} | \psi \rangle =  | \uparrow_\theta \rangle |  \downarrow_\theta \rangle \dots  | \uparrow_\theta \rangle | \downarrow_\theta \rangle  \dots ,  \nonumber \\ 
		& | \uparrow_\theta \rangle = \cos \theta 	| \uparrow \rangle  + 
		i \sin \theta 	| \downarrow \rangle  , 
		| \downarrow_\theta \rangle = \cos \theta 	| \downarrow \rangle  + 
		i \sin \theta 	| \uparrow \rangle , 
	\end{align}
	which is a linear combination of all configurations, including the rarest ones. (At this point, their amplitudes are not in agreement with the infinite-temperature ensemble, but this will be fixed in finite time, as we have seen numerically.) A more pertinent classical analogy would be a stochastic evolution, such as a Ising-Glauber dynamics, in which a rare ``conspiracy'' can lead to an atypical configuration at short time. Then, the tail of the large full distribution will also converge rapidly. 
	
	In summary, the full distribution of an extensive quantity equilibrates as quickly as its low moments, and does not qualify as a potential definition of complexity.

	\subsection{Outcome large deviation in monitored dynamics} \label{sec:measure}
	We now consider another definition of quantum large deviations, using the distribution of measurement outcomes $M_t$ for $t \le t_{\max}$ obtained from a continuous monitoring of an extensive quantity.
	
	This is a generalization of the classical large deviation described in Section~\ref{sec:classical}, which can be viewed as a time distribution of measurement outcomes. Indeed, in classical systems, ideal measurements can be performed in a passive way, gaining information about the system without altering its state. Quantum measurements inevitably perturb the intrinsic unitary dynamics (barring trivial ones that learn nothing from the system), and result in a sequence of random outcomes $M_t$ and an associated quantum trajectory~\cite{Wiseman_Milburn_2009}. Averaging over the outcomes, the system is expected to relax to the infinite-temperature state generically. So, for $t$ sufficiently large, the probability distribution of the outcome $M_t$ will be a large deviation form given by the infinite-temperature ensemble. Then, we expect that a long time sequence of outcomes sample this full distribution with a time-dependent cutoff, as in the classical case (see Section~\ref{sec:classical}). In particular, the large deviation outcomes $M \propto N$ will appear at an exponential time scale $t\sim e^{N}$. 
	
	\begin{figure}
		\centering
		\includegraphics[width=1\linewidth]{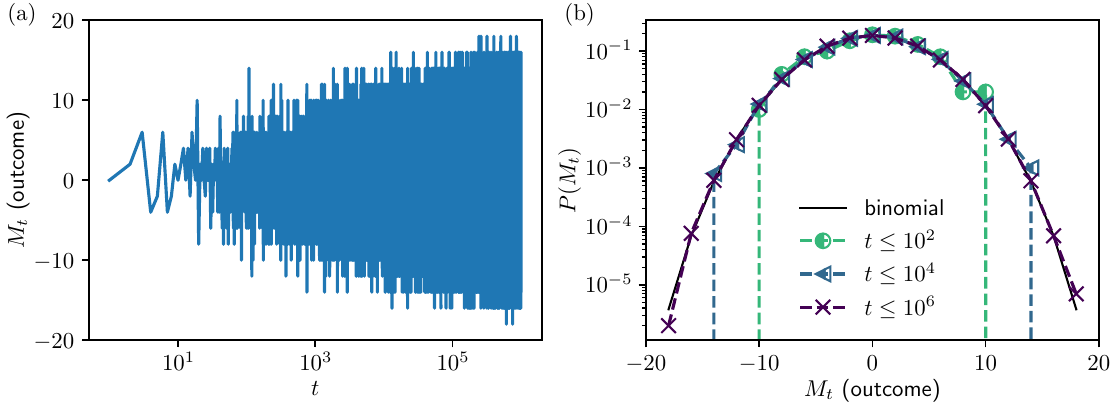}
		\caption{Large deviation of the outcomes time sequence in the kicked Ising model (with $N = 18$ qubits), where the the total magnetization is measured projectively at every timestep. (a) The outcomes time sequence of a single realization,  starting from the N\'eel state $| \psi \rangle = | \uparrow \downarrow \uparrow\downarrow \dots \rangle$. (b) The distribution of the outcomes with $t \le t_{\max}$, $t_{\max} = 10^2, 10^4, 10^6$, compared with the binomial distribution.}
		\label{fig:measure}
	\end{figure}
	In Figure~\ref{fig:measure} we observe the above scenario in the kicked Ising model (see above for definition), where we measure the total magnetization $\mathcal{M} = \sum_j Z_j$ projectively after each unitary evolution time step. In Figure~\ref{fig:measure}-(a), we show the outcome time sequence in a semi-log scale, where it is apparent that more and more atypical outcomes gradually occur at exponentially long time. In Figure~\ref{fig:measure}-(b), the outcome distribution, obtained from a single long quantum trajectory up to $t_{\max}$, is in good agreement with the binomial distribution of the infinite-temperature ensemble modulo a time-dependent cutoff, 
	\begin{align}  \label{eq:PMt}
		P(M_t) :=  \frac{1}{t_{\max}} \sum_{t \le t_{\max}} 
		\delta_{M_t, M}  \sim \begin{cases}
			e^{- N f(M/N)}   &  |M| / N \lesssim m_* \\ 
			0   & \text{otherwise}
		\end{cases}, \quad \text{ where }  f(m_*) = N^{-1} \ln t_{max},
	\end{align}
	and $f$ is the binomial large deviation rate function defined in \eqref{eq:ldf-short} above. Eq.~\eqref{eq:PMt} holds for $t \lesssim 2^N$; when $t \gg 2^N$, the cutoff does not exist anymore as the rarest outcomes have already occurred. 
	
	We remark that the quantum trajectory of the conditioned wavefunction depends on the measurement scheme. In the above numerical example with projective measurement of $\mathcal{M}$, we checked that the wavefunction has volume-law entanglement during most of the quantum trajectory. Now, if one measures all the qubits individually in the computation basis after each timestep, and obtain $M_t$ by summing the outcomes, the wavefunction will remain a matrix product state with $\mathcal{O}(1)$ bond dimension. Nevertheless, the above reasoning does not depend on the measurement scheme, and the outcomes time sequence will have the same behavior as shown in Figure~\ref{fig:measure}. 
	
	One might be concerned that this large deviation does not probe the intrinsic quantum dynamics, which is perturbed by the measurement. In this regard, we may argue that black holes are also monitored by outside observers via their Hawking radiation. Moreover, recent works showed that consistent histories emerge with respect to slowly-evolving extensive quantities in a many-body system, that is, their fluctuation in time can be monitored with vanishing back-reaction on the intrinsic dynamics~\cite{gemmer-16-markov,classicality-thermalization,strasberg-scipost,dhc-prx,cao-pra26}. Finally, continuous monitoring or coupling to external bath is also involved in other probes of quantum chaos, in particular quantum dynamical entropies~\cite{entropy-jorge,dynamical-entropy}. 
	
	

	

	
	

	\subsection{Expectation value large deviation}\label{sec:expectation}
	We now turn to the third kind of quantum large deviation, defined by the distribution of expectation values $ \left< A \right>_t  := \left< \psi(t) |  A | \psi(t) \right>$ over $t \le t_{\max}$~\footnote{More precisely, we may also introduce a fixed early time limit $t_{\min}$ in order to exclude the initial values of the expectation values, which are often atypical. Otherwise, the results below can be easily amended to account for the initial values.},  
	\begin{equation} 
		P_{t_{\max}}(a) :=  \frac1{t_{\max}} \sum_{t \le t_{\max}} \delta( \left< A \right>_t - a ),
	\end{equation}
	or the generation function 
	\begin{equation}
		F_{t_{\max}}(\lambda) :=   \frac1{t_{\max}} \sum_{t \le t_{\max}} e^{\lambda \left< A \right>_t}, 
	\end{equation}
	where $A$ is a Hermitian operator.  Note that $A$ is not necessarily an extensive observable. In fact, we shall first focus on two particular choices of $A$:
	\begin{itemize}
		\item \textbf{Return probability}. By choosing $A = | \psi \rangle \langle \psi |$, the expectation value becomes the return probability 
		\begin{equation}
			\left< A \right>_t =  |\left< \psi | \psi(t) \right>|^2 =: r.
		\end{equation}
		\item \textbf{Spectral form factor}. 
		A further special case of the return probability is the spectral form factor (SFF),
		\begin{equation}
			\mathrm{SFF} :=  \frac1{\mathcal{D}^2}  \left| \mathrm{Tr}[U^t] \right|^2.
		\end{equation}
		Indeed, consider two identical copies of the system with time evolution unitary $U  \otimes U$, and let the initial state be the (infinite-temperature) thermofield double (TFD) defined on two identical copies of the system,
		\begin{equation}
			| \psi \rangle = |\mathrm{TFD} \rangle := \frac1{\sqrt{\mathcal{D}}} \sum_i |i\rangle  | i \rangle
		\end{equation}
		where $|i \rangle, i = 1, \dots, \mathcal{D}$ is an orthonormal basis of the Hilbert space (of one system) and $\mathcal{D}$ is its dimension. Then we have 
		\begin{equation}
			| \langle  \mathrm{TFD} | U \otimes U   |\mathrm{TFD}  \rangle |^2 =  \mathrm{SFF}(2t).  
		\end{equation}
		To remove the factor $2$ above, we may replace the time evolution operator with $U \otimes \mathbf{1} $. The above relation between SFF and TFD extends straightforwardly to finite temperature in a Hamiltonian system.
	\end{itemize}
	We shall consider atypically large values of both quantities. Note that atypically small values (near-zeros) of the spectral form factor, represented by the spikes of $-\ln \mathrm{SFF}$, have been studied in Ref.~\cite{bunin2024fisher,charamis2026quenched}.  
	
	The basic idea is to treat the evolution of the quantum wavefunction in the Hilbert space as a trajectory in the configuration space of a classical system, on which the expectation value is an observable. (As we shall see, the degrees of freedom of the classical system are $\mathcal{D}$ ``XY'' spins.) We then apply the argument of Section~\ref{sec:classical} to predict the evolution of the full distribution. The main point is that the size of the classical system is the Hilbert space dimension $\mathcal{D} \sim e^{N}$, $N$ being the physical size of the quantum system, so that the large deviation time scale is double exponential $\exp(e^{N})$. 
	
	To implement the above idea, we first consider the distribution of $r$ and $\mathrm{SFF}$ in the $t_{\max} \to \infty$ limit for fixed $\mathcal{D}$. For this, let $e^{-i \epsilon_i }, i  = 1, \dots, \mathcal{D}$ be eigenvalues of $U$ ($\epsilon_i$ is the pseudo energy), and define the XY spin variables 
	\begin{equation}
		s_i = s_i(t) = e^{-i \epsilon_i t} \in \mathrm{U}(1) = \{z \in \mathbb{C} : |z| = 1 \} \label{eq:si}.
	\end{equation}
	In terms of these, the return amplitude and SFF can be viewed as Hamiltonians (modulo a pre-factor, see below) of Mattis-like models~\cite{mattis}, 
	\begin{align} \label{eq:rSFFsi}
		r = \sum_{i, j=1}^{\mathcal{D}}  |c_i|^2 |c_j|^2 s_i s_j^* ,   \quad 
		\mathrm{SFF} = \frac{1}{\mathcal{D}^2} \sum_{i, j=1}^{\mathcal{D}}  s_i s_j^*, \text{ where }  c_i :=  \langle \epsilon_i | \psi \rangle, 
	\end{align}
	and $| \epsilon_i \rangle $ is the $i$-th pseudo eigenstate. 
	
\begin{figure}
		\centering
		\includegraphics[width=1\linewidth]{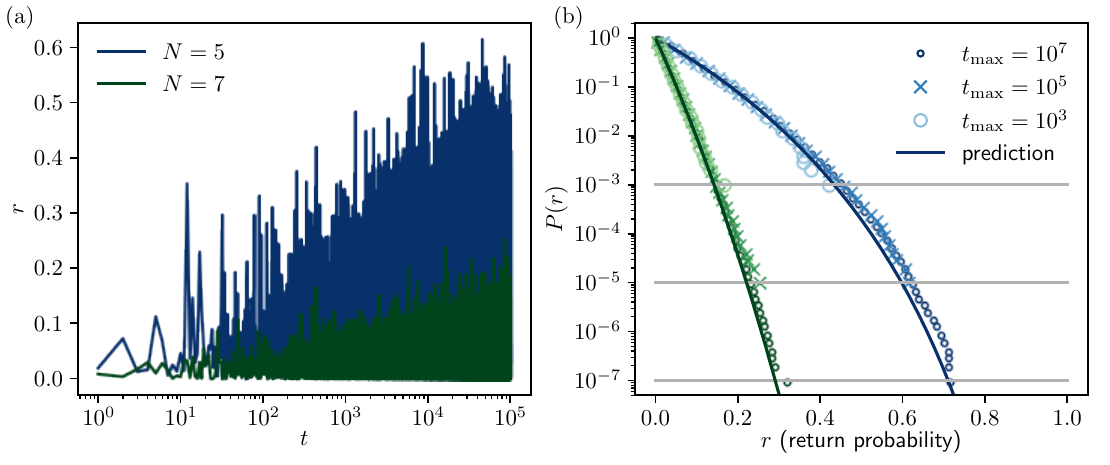}
		\caption{Return probability in the kicked chaotic Ising model of size $N = 5$ and $N= 7$ in the longtime regime, with initial state $|\psi \rangle =  | \uparrow \uparrow\dots \uparrow \rangle$. (a) Return probability as function of time. (b) The time distribution $P_{t_{\max}}$ for $t_{\max} = 10^3, 10^5, 10^7$ compared to prediction~\eqref{eq:Ptmax-quantum}, and the solution of the Mattis model~\eqref{eq:Pinfty-S}-\eqref{eq:relations}. }
		\label{fig:long-return}
	\end{figure}

\begin{figure}
		\centering
		\includegraphics[width=1\linewidth]{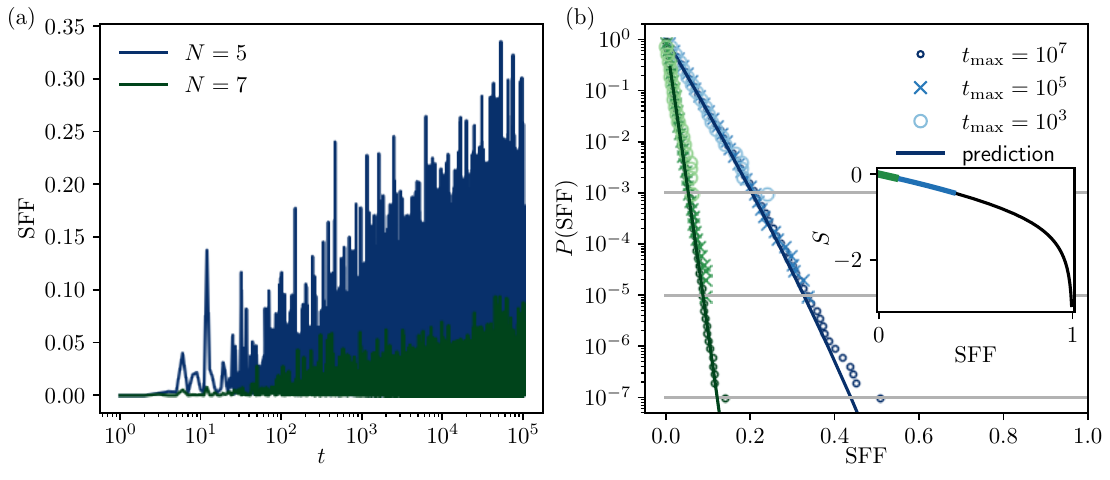}
		\caption{Spectral form factor (SFF) in the kicked chaotic Ising model of size $N = 5$ and $N= 7$ in the longtime regime, with initial state $|\psi \rangle =  | \uparrow \uparrow\dots \uparrow \rangle$. (a) SFF as function of time. (b) The time distribution $P_{t_{\max}}$ for $t_{\max} = 10^3, 10^5, 10^7$ compared to prediction~\eqref{eq:Ptmax-quantum}, and the solution of the Mattis model~\eqref{eq:Pinfty-S}-\eqref{eq:relations}. The inset plots the entropy density and shows the regions explored numerically.}
		\label{fig:long-sff}
	\end{figure}

	In the long time limit, for a generic $U$, the XY spin variables $s_i = s_i(t)$ sample uniformly the configuration space $\mathrm{U}(1)^{\mathcal{D}}$: this torus plays the r\^ole usually played by the energy shell in classical mechanics. Therefore, the $t_{\max} = \infty$ distribution of $r$ and $\mathrm{SFF}$ can be computed using equilibrium statistical mechanics tools. We introduce the partition function of the Mattis model, 
	\begin{equation} \label{eq:Z-mattis}
		Z(\beta) := \mathbb{E} [e^{ -\beta \mathcal{D} h}], \text{ where } \mathbb{E}[\dots] := \int_{|s_i| \in \mathrm{U}(1)} \prod_{i=1}^{\mathcal{D}} \frac{\mathrm{d} s_i}{2\pi i s_i}  [\dots]
	\end{equation}
	is the average over the the configuration space, and $-h = r$ or $\mathrm{SFF}$. because of this normalization, the entropy is zero for the most typical configurations. Then the energy distribution has a large deviation form, whose rate function is given by the entropy density (up to a minus sign),
	\begin{equation} \label{eq:Pinfty-S}
		P_\infty(a) \sim e^{ \mathcal{D} S(-a) }, \; \text{where } a = r \text{ or } \mathrm{SFF}. 
	\end{equation} 
	Now, the entropy density can be computed exactly in the $\mathcal{D} \to \infty$ limit, provided the assumption that the $|c_i|$'s are Gaussians with zero mean and $1 / \mathcal{D}$ variance in the return probability case (this is a valid assumption for the kicked Ising model for a initial state such as $|\psi \rangle = | \uparrow \uparrow\dots \uparrow \rangle$, as we have checked above).  Referring to Appendix \ref{sec:mattis} for calculation details, we state the result here:
	\begin{equation} \label{eq:entropy}
		S(-a) = - 2 \beta^{-1} |z|_*^2 +  \int p(v) \ln I_0(2 v^2 |z|_*)  \mathrm{d} v, 
	\end{equation}
	where $I_n$ is the modified Bessel function of the first kind, 
	\begin{equation} \label{eq:pv}
		p(v) = 
		\begin{cases}
			e^{-v^2 / 2} / \sqrt{2 \pi}  & \text{for return probability} \\ 
			\delta(v - 1)  & \text{for SFF},
		\end{cases}
	\end{equation}
	and $\beta$ and $z_*$ are related to $a$ by 
	\begin{equation} \label{eq:relations}
		a = \beta^{-2} |z|_*^2, \;  |z|_* = \beta \int p(v) v^2 \frac{I_1 (2 v^2 |z|_*)}{I_0 (2 v^2 |z|_*)}  \mathrm{d} v . 
	\end{equation}
	See Figures~\ref{fig:long-return} and \ref{fig:long-sff} for plots of the analytical prediction of $ P_\infty(a) $.
	
	Having determined the distribution of $r$ and $\mathrm{SFF}$ in the longtime limit as a large deviation form, we can apply the argument of Section~\ref{sec:classical}, which predicts that, for large but finite $t_{\max}$, the distribution of expectation values is given by infinite-time large deviation form~\eqref{eq:Pinfty-S} with a sliding cut-off, 
	\begin{equation} \label{eq:Ptmax-quantum}
		P_{t_{\max}}(a) \sim \begin{cases}
			e^{ \mathcal{D} S(-a) }  &  a \lesssim a_* \\ 
			0  & a \gtrsim a_*
		\end{cases}, \text{ where }   S(-a_*) =  - \mathcal{D}^{-1} \ln t_{\max}.
	\end{equation} 
	There is only one cutoff $a_*$ because $a \ge 0$ and the most typical value is $a = 0$ (more precisely $a \ll 1$). In general, the cutoff disappears when $ \mathcal{D} \ln t \gtrsim  \max S - \min S$, when the rarest values are visited.  Here, $S(-a) \sim \ln (1 - a)$ as $a \to 1$ is unbounded from below (see Appendix~\ref{sec:mattis}), so the cutoff always exists and  approaches $1$ as $t_{\max} \to \infty$.
	
	We tested the above predictions using longtime numerical simulation of the kicked Ising model~\eqref{eq:kicked-def} of rather small sizes. The results are plotted in Figure~\ref{fig:long-return} for return probability and Figure~\ref{fig:long-sff} for SFF. Both quantities evolve rapidly from the initial value to a typical one where the entropy is maximum. Then, gradually, more atpyical regions are visited; these excursions are the spikes in Figure~\ref{fig:long-return}-(a) and Figure~\ref{fig:long-sff}-(a). Larger spikes are found at larger times. In Figure~\ref{fig:long-return}-(b) and Figure~\ref{fig:long-sff}-(b), we find an excellent agreement between $P_{t_{\max}}$ obtained numerically and the Ansatz~\eqref{eq:Ptmax-quantum} combined with analytical predictions~\eqref{eq:Pinfty-S}-\eqref{eq:relations} from the statistical mechanics of the Mattis model. 
	
	As the cutoff value $a_*$ in \eqref{eq:Ptmax-quantum} evolves extremely slowly, with a double exponential time scale $t \sim e^{\mathcal{D}} \sim \exp({e^{N}})$, it provides a notion of complexity in quantum dynamics. The complexity is reflected in the fact that, at very longtime, the seemingly featureless quantum evolution contains brief excursions to more and more atypical corners of the Hilbert space. In principle, such excursions can be observed with a high probability, provided that the system can maintain coherence for sufficiently long time. 

	So far we focused on the return probability and the spectral form factor. Clearly, the analogy with a classical XY model can be formulated for other quantities as well, by writing them in terms of the variables $s_i$~\eqref{eq:si}. Then we may obtain similar prediction on the large deviations, as we shall do for a few quantities in what follows.
	
	\paragraph{Expectation value.} For a general Hermitian operator $A$ and an initial state $| \psi \rangle$, we have
	\begin{equation}
		\left<\psi(t) | A | \psi(t)  \right> = \sum_{i,j=1}^{\mathcal{D}} J_{ji} s_j^*  s_i ,  \quad \text{where } J_{ji} := \langle \epsilon_j | A | \epsilon_i \rangle \langle \epsilon_i | \psi \rangle  
		\langle \psi |  \epsilon_j  \rangle . 
	\end{equation}
	We can further split the sum into the diagonal and off-diagonal parts, 
	\begin{equation}
		\left<\psi(t) | A | \psi(t)  \right>  = \sum_{i} J_{ii} + 
		\sum_{i\ne j }^{\mathcal{D}} J_{ji} s_j^* s_i  . 
	\end{equation}
	The first term is time-independent and does not interest us here. The couplings $J_{ji}$ in the second term may be estimated by the eigenstate thermalization hypothesis (ETH) \cite{Srednicki1994,DAlessio2016,FoiniKurchan2019}. For this, assume that $\mathcal{D}^{-1} \mathrm{Tr}[A^2] \sim \mathcal{O}(1)$ (this is the case for a local operator such as $Z_i$ in a spin chain). Then, in the Floquet case with no conservation law, $ \langle \epsilon_j | A | \epsilon_i \rangle$ resembles a full random matrix whose elements have zero mean and variance $1 / \mathcal{D}$.
	Recalling also that $ |\langle \epsilon_i | \psi \rangle| \sim 1/\sqrt{D}$, we have
	\begin{equation} \label{eq:Jij}
		|J_{ji}| \sim \mathcal{D}^{-3/2}, \quad i \ne j. 
	\end{equation}
	With this scaling, the time-dependent part of $\left<\psi(t) | A | \psi(t)  \right>$ equals the energy density of a XY spin-glass~\cite{GabayToulouse1981} with $\mathcal{D}$ spins. Its typical values are of order $1 / \sqrt{\mathcal{D}}$ and atypical values are of order $1$ will occur at $ t\sim e^{\mathcal{D}} \sim \exp(e^{N})$. 
	
\begin{figure}
		\centering
		\includegraphics[width=1\linewidth]{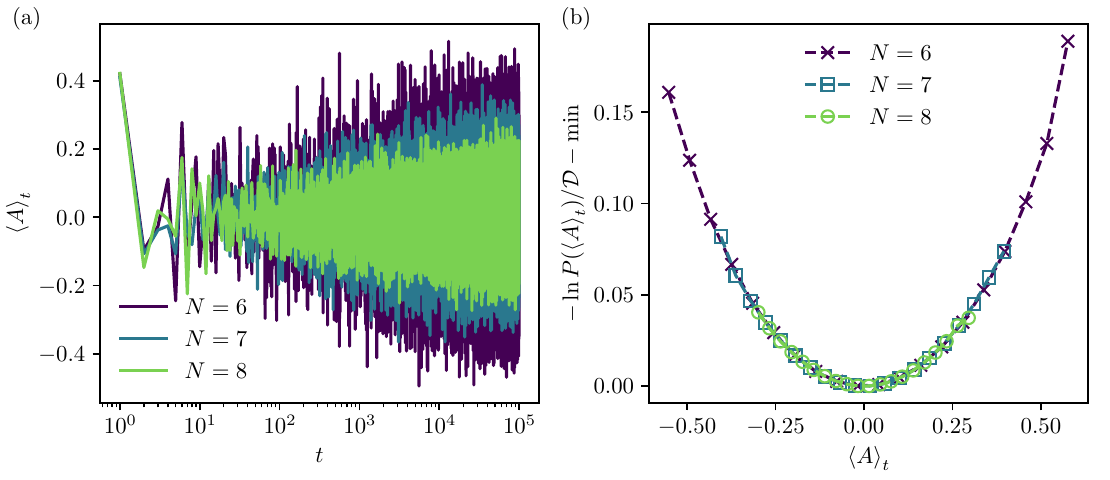}
		\caption{Rare fluctuations of the expectation value $\left< A \right>_t = \left<\psi(t) | A | \psi(t)  \right>$ in the kicked Ising model~\eqref{eq:kicked-def} of size $N = 6, 7, 8$, with $| \psi \rangle = | \uparrow  \uparrow \dots \uparrow\rangle$ and $A = Z_{\lfloor N / 2 \rfloor}$. (a) Time evolution of the expectation value up to $t = 10^5$. (b) We take the histogram of the expectation values for $t \in [10, 10^6]$, take the log, divide by $\mathcal{D} = 2^N$, and subtract the maximum value. The resulting large deviation rate for different $N$ agrees with each other, modulo a cutoff.}
		\label{fig:At}
	\end{figure}

	In Figure~\ref{fig:At} we study numerically the expectation value large deviations in the kicked Ising model~\eqref{eq:kicked-def}. In Figure~\ref{fig:At}-(a), we see that the expectation value has again larger and larger peaks (visible in log-linear scale), following an initial decay. Although we do not solve the corresponding XY model analytically, we numerically extract its energy large deviation rate function as probed by the time evolution, see caption of Figure~\ref{fig:At} for details. In Figure~\ref{fig:At}-(b), we observe that the extracted large deviation function does not depend on $N$, which validates the large deviation principle. We also observe the sharp cutoffs due to finite $t_{\max}$.  
	
	In a Hamiltonian system with energy conservation, the matrix elements have more structure. Assuming that the initial state has an sub-extensive energy fluctuation around $\bar E$, ETH states that 
	\begin{equation}
		\langle \epsilon_j | A | \epsilon_i \rangle e^{- N s(\bar E / N)/2} f_A(\bar E,\omega)\, R_{ij}, i \ne j.
	\end{equation}
	where $s(\varepsilon)$ is the entropy density (as a function of the energy density)
	\begin{equation}
		\bar E=\frac{E_i+E_j}{2}, \omega=E_i-E_j.
	\end{equation}
	and The $R_{ij}$ are random variables of variance one. We have $ \langle \epsilon_j | A | \epsilon_i \rangle \sim 1 / \sqrt{\mathcal{D}'}$, where we denoted $ \mathcal{D}' := e^{ N s(\bar E / N) }$ Meanwhile, $\left< \epsilon_i | \psi \right> \sim 1/\sqrt{\mathcal{D}'}$ for $\sim \mathcal{D}'$ eigenstates. In summary, we \eqref{eq:Jij} still holds with $\mathcal{D}$ effectively replaced by $\mathcal{D}'$. So, as long as $| \psi \rangle $ relaxes  to a thermal state with nonzero entropy density, we still have an double exponential large deviation time $t \sim e^{\mathcal{D}'}$. 
	
\begin{figure}
		\centering 
		\includegraphics[width=1\linewidth]{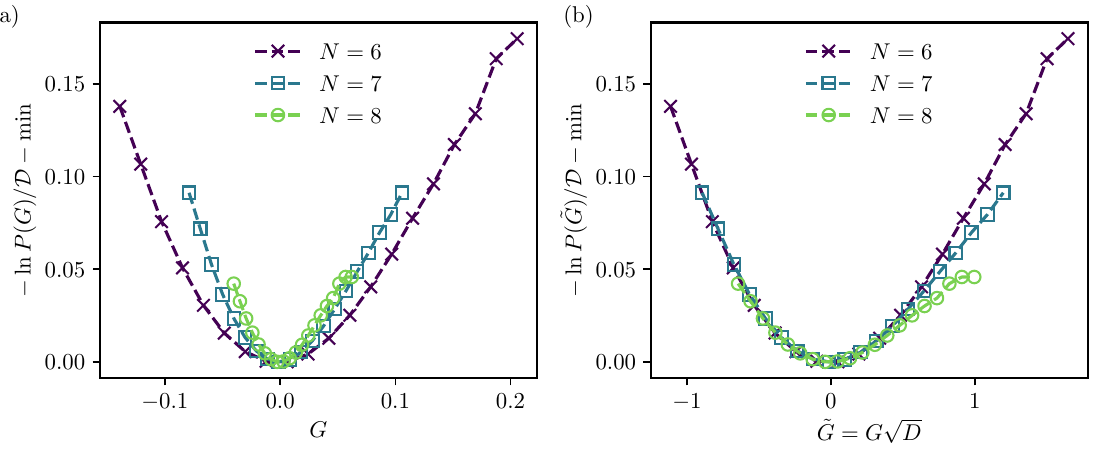}
		\caption{Rare fluctuations of the auto-correlation function for the kicked Ising model~\eqref{eq:kicked-def} of size $N = 6, 7, 8$ and $A = Z_{\lfloor N / 2 \rfloor}$. (a) Numerical large deviation rate extracted with the same method as Figure~\ref{fig:At}. (b) Same as (a), but with the rescaled variable $\tilde{G} = G \sqrt{D}$.}
		\label{fig:Gt}
	\end{figure}

	\paragraph{Auto-correlation function.} For a general Hermitian operator $A$, the auto-correlation function (Green function) is  
	\begin{equation}
		G(t) := \mathcal{D}^{-1} \mathrm{Tr}[ A(t) A ] = \sum_{i,j=1}^{\mathcal{D}} K_{ji} s_j^* s_i  , \; K_{ji} =  \mathcal{D}^{-1} | \langle \epsilon_j | A | \epsilon_i \rangle |^2  
	\end{equation}
	With the same assumption on $A$ as above, the coupling constants scale as 
	\begin{equation}
		K_{ji}  \sim \mathcal{D}^{-2},  \quad i \ne j.
	\end{equation}
	Note that $K_{ji} \mathcal{D} \sim 1/\mathcal{D}$ is the correct scaling of an all-to-all ferromagnetic XY model. So, rare \textit{positive} fluctuations of $G(t)$ correspond to atypical values of the energy density realized by ferromagnetic configurations. Meanwhile, rare \textit{negative} fluctuations of $G(t)$ cannot stem from ferromagnetism but only from the disorder involved in $K_{ji}$, which is ``incorrectly'' normalized, such that $ \sqrt{D} G$ is the energy density. As a result, we expect that the rare fluctuations are governed by the following large deviation form in the $\mathcal{D} \to\infty$ limit
	\begin{equation} \label{eq:PG-G}
		P_\infty(G) \sim \begin{cases}
			\exp( \mathcal{D} S(G ) )  & G > 0 \\
			\exp( \mathcal{D} S(G \sqrt{D}) ) & G < 0
		\end{cases}. 
	\end{equation}
	In other words, positive fluctuations are parametrically larger than negative ones, while both have a $e^{\mathcal{D}}$ time scale. We tested this Ansatz numerically by extracting the large deviation form using both $G$ and $ \sqrt{D} G$ as energy density. In Figure~\ref{fig:Gt}-(a), we see a reasonable agreement with \eqref{eq:PG-G} for $G > 0$ (it improves as $N$ increases), and in Figure~\ref{fig:Gt}-(b), a good agreement for $G < 0$ is found. It remains to understand how this analysis matches the one in Ref. \cite{bouverot2025random}, which is based on the connection with random matrix theory.

	\paragraph{Operator size.} Finally we briefly consider fluctuations of the mean operator size, which can be viewed as the expectation value of a super-operator $\mathbf{S}$ that acts on the doubled Hilbert space of operators. (In this respect, the auto-correlation function above is a return amplitude in the operator space.) In a qubit system, Pauli strings, that is, products of Pauli operators (including identity) form an orthonormal basis of the operator space, endowed with the inner product 
	\begin{equation}
		(A | B) := \mathcal{D}^{-1} \mathrm{Tr}[A^\dagger B].
	\end{equation}
	The size of a Pauli string, $s(P)$, is the number of non-identity factors. Then, the operator size super-operator is defined by
	\begin{equation}
		\mathbf{S} | P ) :=  s(P) | P ),
	\end{equation}
	For an operator $A$ such that $(A | A) = 1$, The mean size of an $A$ under Heisenberg-picture evolution is 
	\begin{equation}
		\bar{\mathbf{S}}(t) :=  ( A(t) |  \mathbf{S} | A(t) ),
	\end{equation}
	In terms of the XY spins, we have a model with quartic interaction
	\begin{equation}
		\bar{\mathbf{S}}(t) = 
		\sum_{ijk\ell}  K_{ijk\ell}  s_{i} s_j^* s_k^* s_\ell, \quad 
		K_{ijk\ell} := \langle \epsilon_j | A | \epsilon_i \rangle \langle \epsilon_\ell | A | \epsilon_k \rangle^*  \, (k \ell |\mathbf{S} | i j ) 
	\end{equation}
	where $ (k \ell |\mathbf{S} | i j ) $ is the operator size super-operator in the doubled pseudo-energy basis. Now, a typical operator (taken randomly in the operator Hilbert space) has an average size close to $ 3N / 4$, because there are $3$ non-identity Pauli's out of $4$. We thus expect that, for any initial operator $A$ (in particular, one of small size), $\bar{\mathbf{S}}(t) $ will rapidly grow to $3 N / 4$. Then, rare fluctuations far away from the typical value will occur at $t\sim \exp(e^{N})$. 

	\section{Discussion}
	We have studied large deviations in many-body quantum dynamics defined in three ways, with distinct physical meanings and time scales. \textit{Potential} (but not actually observed) rare fluctuations, encoded in the full distribution, become present first, much earlier than in a classical deterministic system, thanks to quantum parallelism. Observed rare fluctuations, in the sense of atypical measurement outcomes under continuous monitoring, occur at $t \sim e^{N}$, as in classical systems. Finally, at the quantum recurrence time scale $ t \sim \exp( e^{N})$, we encounter expectation value large spikes, that is, brief moments where  \textit{one-shot} observation of rare fluctuations occurs with $\mathcal{O}(1)$ probability. 
	
	The time distribution of both outcomes and expectation values has a large deviation form with slowly evolving cutoffs, which we propose as measures of quantum complexity. Namely, we characterize \textit{complexity as the possibility of extraordinary}. The expectation-based complexity characterizes the extent to which a coherently evolving wavefunction (or operator) exhaustively explores the Hilbert/operator space, in particular its atypical corners. Its saturation time scale $\exp(e^{N})$ is much larger than that of Krylov and Nielsen complexity, although the latter also diverges with the precision $\epsilon$.  Meanwhile, the measurement-based complexity is about an open system displaying more and more atypical phenomena to observers, and has a $e^{N}$ saturation time scale, identical to that of classical large deviation, and of Krylov/Nielsen complexity.
	
	The cutoff values in our proposed complexities depend clearly on the operator; in particular, they scale with the operator, and their time evolution is also dictated by the distribution that is being sampled. A more choice-independent quantity is the number $M$ of effectively independent samples (of outcomes or expectation values) realized by $t = t_{\max}$. Its logarithm, related the cutoffs by the ``second principle'' arguments above, is given by
	$$   \ln M  \sim \ln (t_{\max} / t_c) = \ln t_{\max} - \ln t_c, $$
	where $t_c$ is a correlation time. For generic operators and measurement schemes in a chaotic system, $t_c$ is a rather short relaxation time that scales at most as a power-law of the system. So for $t_{\max} \sim N$ or $\sim \mathcal{D}$, which is our time scales of interest, $\ln t_c \ll \ln t_{\max}$ and affects $\ln M$ by a negligible relative amount (in fact, we have set $t_c = 1$ in our numerical tests). In this sense, the growth of our complexities depends weakly on the choice of operators. 
	
	It will be interesting to see to what extent obstructions to thermalization (such as integrability) affect the growth of the large-deviation complexities. Also, in the holography context, we wonder what are the gravity duals of the proposed complexities.    
	
	
	
	\section*{Acknowledgments}
	
	We wish to thank Hugo Camargo, Laura Foini, Shiraz Minwalla, Adam Nahum, Zohar Nussinov and Giuseppe Policastro for useful discussions. 
	
	\appendix
	\section{Statistical mechanics of the Mattis model} \label{sec:mattis}
	In this Appendix we sketch the solution of the Mattis model defined by the partition function~\eqref{eq:Z-mattis}, which we rewrite as follows,
	\begin{equation}
		Z = \mathbb{E} \left[e^{ \frac{\beta}{\mathcal{D}} \sum_{i,j=1}^{\mathcal{D}} v_i^2 v_j^2 s_i s_j^* } \right]
	\end{equation}
	where $v_i \equiv 1$ for the SFF and $v_i \sim \mathcal{N}(0, 1)$ for the return probability. We then apply the Hubbard–Stratonovich decoupling by introducing $z \in \mathbf{C}$, 
	\begin{equation}
		Z =  \int \frac{\mathrm{d}^2 z}{\pi} \mathbb{E}\left[
		e^{- \beta^{-1} \mathcal{D} |z|^2 + \sum_i v_i^2 (z s_i^* + s_i z^*) }
		\right],
	\end{equation}
	and perform the integral over the $s_i$'s. Replace the $\sum_i F(v_i)$ by integral $ \mathcal{D} \int \mathrm{d} v  p(v) F(v)$, where $p$ is defined in \eqref{eq:pv} (this is exact in the large $\mathcal{D}$ limit) we obtain an effective action,  
	\begin{equation}
		Z =  \int \frac{\mathrm{d}^2 z}{\pi} e^{-  \mathcal{D} L(z) }, \, 
		L(z) = \beta^{-1} |z|^2 - \int \mathrm{d} v p(v) \ln I_0(2 v^2 |z|) .
	\end{equation}
	It follows that the free energy density is 
	\begin{equation} 
		\beta f(\beta) := - \lim_{\mathcal{D} \to \infty} \ln Z(\beta) = 
		\beta^{-1} |z|_*^2 - \int \mathrm{d} v p(v) \ln I_0(2 v^2 |z|_*) ,
	\end{equation}
	where $|z|_*$ is the minimum locus of $L(z)$, given by \eqref{eq:relations}, which we copy here,
	\begin{equation} \label{eq:zstar}
		|z|_* = \beta \int p(v) v^2 \frac{I_1 (2 v^2 |z|_*)}{I_0 (2 v^2 |z|_*)}  \mathrm{d} v . 
	\end{equation}
	Then the entropy density is given by standard thermodynamics relations,
	\begin{equation} \label{eq:Sa-a-app}
		S(-a) = - \beta (a + f(\beta) ), \, -a = \partial_\beta (\beta f) = -\beta^{-2} |z|_*^2. 
	\end{equation} 
	Note that the point $a = 0$ corresponds to $\beta = \beta_*$ where $\beta_* = 1/3$ (for return probability) and $\beta_* = 1$ (for SFF). $\beta_*$ is a critical point of the model.  $a = 0$ throughout the high temperature phase, and $a \ne 0$ in the low-temperature phase. In the zero temperature limit, we have $|z|_* \gg 1$, \eqref{eq:zstar} is approximated as 
	\begin{equation}
		\beta^{-1} |z|_* =  \int p(v) \left(v^2 - \frac1{4 |z|_*}  \right)   \mathrm{d} v  + \mathcal{O}(|z|_*^{-2}) = 1 - \frac1{4 |z|_*} + \mathcal{O}(|z|_*^{-2})
	\end{equation}
	for both choices of $p(v)$ in \eqref{eq:pv}. Therefore, by \eqref{eq:Sa-a-app} we have 
	\begin{equation}
		1 - \sqrt{a} = \frac1{4 |z|_*} + \mathcal{O}(|z|_*^{-2}), \beta = |z|_* + \mathcal{O}(1).
	\end{equation}
	Near $a = 1$, we have thus
	\begin{equation}
		S(-a) = - \int \beta \mathrm{d} a \sim - \int  - 2 |z|_* \mathrm{d} (1 - \sqrt{a}) \sim 
		\int \frac{\mathrm{d} |z|_*}{2 |z_*|} \sim \frac12 \ln (1 - a). 
	\end{equation}

	\bibliography{paper.bib}
\end{document}